\documentclass[namedreferences]{SolarPhysics}
\usepackage[optionalrh]{spr-sola-addons} 
\usepackage{graphicx}                    
\usepackage{color}                       
\usepackage{url}                         

\newcommand{\apropto}{\;
  \raise0.3ex\hbox{$\propto$\kern-0.75em\raise-1.1ex\hbox{$\sim$
  }}\;\hskip-2pt }

\begin{document}

\begin{article}

\begin{opening}

\title{Solar Grand Minima and Random Fluctuations in Dynamo Parameters}

%
\author{D.~\surname{Moss}$^{1}$\sep
        D.~\surname{Sokoloff}$^{2}$\sep
        I.~\surname{Usoskin}$^{3}$\sep
        V.~\surname{Tutubalin}$^{4}$
       }

%
\runningauthor{Moss et al.}
\runningtitle{Solar Grand Minima and random fluctuations in dynamo parameters}

%
\institute{$^{1}$ School of Mathematics, University of Manchester, Manchester M13 9PL, UK.
                     email: \url{moss@ma.man.ac.uk} \\
             $^{2}$ Department of Physics, Moscow State University, Moscow, 119992, Russia.
                     email: \url{sokoloff@dds.srcc.msu.su} \\
             $^{3}$ Sodankyl\"a Geophysical Observatory (Oulu Unit), 90014 University of Oulu, Finland.
                     email: \url{Ilya.Usoskin@oulu.fi}\\
             $^{4}$ Department of Mechanics and Mathematics, Moscow State University, Moscow, 119992, Russia.
                     email: \url{vntutubalin@yandex.ru}\\
             }

\begin{abstract}
We consider to what extent the long-term dynamics of cyclic solar
activity in the form of Grand Minima can be associated with random
fluctuations of the parameters governing the solar dynamo. We
consider fluctuations of the alpha-coefficient in the conventional
Parker migratory dynamo, and also in slightly more sophisticated
dynamo models, and demonstrate that they can mimic the gross
features of the phenomenon of the occurrence of Grand Minima over a
suitable parameter range. The temporal distribution of these Grand
Minima appears chaotic, with a more or less exponential waiting time
distribution, typical of Poisson processes. In contrast however, the
available reconstruction of Grand Minima statistics based on
cosmogenic isotope data demonstrates substantial deviations
from this exponential law. We were unable to reproduce the
non-Poissonic tail of the waiting time distribution either in the
framework of a simple alpha-quenched Parker model, or in its
straightforward generalization, nor in simple models with feedback
on the differential rotation. We suggest that the disagreement may
only be apparent and is plausibly related to the limited
observational data, and that the observations
 and results of numerical modeling can be consistent and
represent physically similar dynamo regimes.

\end{abstract}

%

\end{opening}

%

%

\section{Introduction}

The solar cycle is believed to be a result of dynamo action
occurring somewhere inside the solar convective envelope. According
to the classical Parker (1955) model, this dynamo action can be
envisaged as follows. Differential rotation $\Omega$ produces
toroidal magnetic field from poloidal, while the "$\alpha$-effect"
associated with the helicity of the velocity field produces poloidal
magnetic field from toroidal. According to this scheme, the solar
cycle length is identified with the dynamo time-scale, which can be
estimated from the product of the amplitudes of the $\alpha$-effect
and rotational shear ($\partial \Omega / \partial r$, $r$ being the
radial coordinate), appropriately normalized with the turbulent
diffusion coefficient (these yielding the dimensionless dynamo
number), and with the turbulent diffusion time. The Parker model
results in a periodic process in the form of propagation of a
toroidal field pattern in the latitudinal direction (the "butterfly
diagram"). A suitable choice of governing dynamo parameters gives
equatorward pattern propagation as well as allowing the cycle period
to agree with observations. More realistic dynamo models try to
demonstrate that some plausible choice of parameters is compatible
with available observational information from, say, helioseismology,
or elaborates this simple scheme by various additional details, such
as meridional circulation (see Petrovay, 2000;
Charbonneau, 2005, for reviews).

In fact, the solar cycle is far from being a strictly periodic
phenomenon. The amplitude of solar cycles varies substantially in
time and reaches unusually large levels during the so-called Grand
Maxima, one of which is now believed to be occurring. From
time to time the level of solar cyclic activity becomes extremely
low if not disappearing completely. Such minima of the cyclic activity
are known as Grand Minima, the most well-known example being the
Maunder Minimum, which occurred in the middle 17-th - beginning of
18-th centuries. The statistics of Grand Minima (and Maxima) can be
to some extent reconstructed from data  on cosmogenic
 isotope $^{14}$C in tree rings (Usoskin {\it et al.}, 2007).
Quantification of the sequence of such events is still a
contentious topic.  It is important that the isotopic data provides
a much longer record of Grand Minima/Maxima than do the sunspot
observations. Moreover, the sequence of Grand Minima (and/or Maxima)
appears to be random, rather  than a periodic process.

It is known that simple deterministic numerical dynamo models of the
solar cycle, which essentially develop the ideas of the Parker
migratory dynamo, can give events comparable with  Grand
Minima/Maxima ({\it e.g.} Brandenburg {\it et al.}, 1989), even showing behaviour
which is irregular
and chaotic in time (see {\it e.g.} Jennings and Weiss, 1991; Jennings, 1991;
Tobias {\it et al.}, 1995; Covas {\it et al.}, 1998  -- see also Moss and Brooke, 2000
in a more complex model). The presence of a long-term
dynamics needs however an explanation. The most straightforward idea
here is to recognize that the $\alpha$-effect, being the result of
the electromotive force averaged over turbulent vortices, can
contain a fluctuating contribution (Hoyng, 1993; Hoyng {\it et al.}, 1994;
Ossendrijver and Hoyng, 1996).  The idea can lead to events similar to
the Maunder Minimum on the timescale of centuries (see {\it e.g.}
Tworkowski {\it et al.}, 1998; also  Brandenburg and Spiegel, 2008).

The aim of this paper is to investigate the long-term dynamics of
solar  activity by confronting the predictions of a Parker migratory
dynamo model containing a random contribution to the
$\alpha$-coefficient with the available data concerning the sequence
of Grand Minima and Maxima, as inferred from the isotopic data. We
also consider, more briefly, a more sophisticated dynamo model. Our
general conclusion is that the fluctuations in  the dynamo governing
parameters can lead to phenomena similar to the Grand Maxima and
Minima, in that the temporal distribution of the events appears
chaotic.

We recognize a disagreement between the observational data and
numerical simulations of our dynamo model, in that the statistics of
the waiting time distributions of Grand Minima appear to have
exponential tails, in contrast to the isotopic data  in which the
temporal distribution of Grand Minima and Maxima demonstrate a
substantial deviation from exponential statistics. We argue that the
disagreement plausibly is only apparent, and is connected with the
limited extent in time of the observations, and that observations
and modeling may represent physically similar dynamo regimes.

In this paper we concentrate on the distribution of Grand Minima.

\section{Long-term Dynamics of Solar Activity from the Isotopic Data}

Our intention is to  test whether a simple physical model can
reproduce the basic phenomena of the long-term solar dynamics. Of
course, a detailed explanation of the phenomena needs a much more
realistic model, including at least a 2D description of the solar
magnetic field, realistic solar rotation curve, etc. Moreover, we do
not exclude {\it a priori} that the phenomena could have some
alternative explanation. We base our initial analysis on a simple
illustrative model, rather than on something more realistic, in order to
isolate physical phenomena and to take into account the quite
limited status of the actual observational information.

We focus our attention on the long-term dynamics of solar activity on
the time-scale of centuries. We note however that the idea of random
fluctuations of the dynamo governing parameters can be instructive
in explaining the stochastic features of short-term dynamics of solar
activity, on the timescale of a few solar cycles ({\it e.g.} Moss {\it et al.}
1992; Hoyng {\it et al.}, 1994).

We summarize the properties of the long-term solar dynamics based on
the analysis of the data on cosmogenic isotope $^{14}$C in
tree-rings performed by Usoskin {\it et al.} (2007) for the Holocene (the
last 11,000 years or about 1,000 solar cycles). From time to time,
the cyclic solar activity as deduced from this proxy data
demonstrates phenomena resembling Grand Maxima and Grand Minima.
Usoskin {\it et al.} (2007) identified 27 Grand Minima of total duration of 1880
years (or about 17\% of  the entire period) during the 1,000 solar
cycles\footnote{Even though this data set is a reconstruction and
may contain uncertainties, we will regard and refer to it as the
"real" solar activity series throughout the paper, to emphasize its
difference from the purely synthetic data modeled here.}.

\begin{figure}
\begin{center}
\includegraphics[width=0.45\textwidth]{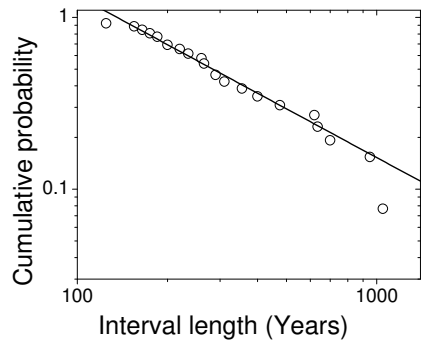}
\end{center}
\caption[]{\label{WTD} Waiting time distribution between the
 subsequent Grand Minima (see Usoskin {\it et al.}, 2007) together
 with the best fit  power law.
 The first and the last point were excluded from the fitting.}
\end{figure}

The
sequence of Grand Maxima/Minima appears chaotic\footnote{We do not
claim here that this sequence {\it is} random in any mathematically
rigorous sense, as proof of such a statement would require many more
proxy Grand Minima/Maxima events than are available.}. An important
parameter characterizing intrinsic features of
quasi-chaotic/stochastic
 processes is the distribution of waiting times (hereafter WTD)
 between successive Grand Minima.
A simple Poisson process ({\it i.e.} with the probability of occurrence of Grand
Minimum being constant and independent of the previous
history of the system) is characterized by an exponential WTD.
On the other hand, significant deviation from an exponential
tail in the  WTD ({\it e.g.}, such as a power law) implies an essentially
non-Poisson process ({\it e.g.} self-organized criticality, or
accumulation and release of energy). The WTD for solar Grand Minima
identified via cosmogenic isotopes by Usoskin {\it et al.} (2007) displays
significant deviation from exponential form (see Figure~\ref{WTD}).

\begin{figure}
\begin{center}
\includegraphics[width=0.45\textwidth]{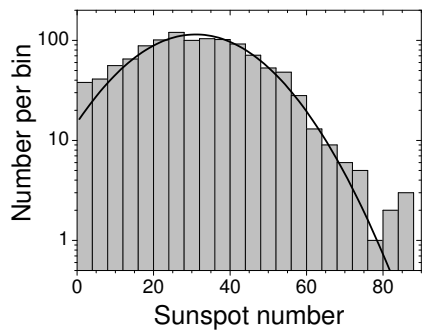}
\end{center}
\caption[]{\label{FD} Histogram (the distribution function) of the
decadal solar activity (quantified in sunspot number)
 during the Holocene (see Usoskin {\it et al.}, 2007) together
 with the best fit normal distribution.}
\end{figure}

In the forthcoming sections we try to reproduce, in  a statistical
sense, the appearance of Grand Minima in the framework of different
dynamo models. As criteria we will consider the shape of the
distribution of waiting times between subsequent Grand Minima
(exponential or power law) and the distribution function (DF) of the
solar activity (assumed to correlate with sunspot number (SN) -- see
Figure~\ref{FD}.)


\section{The Dynamo Model}
\subsection{The Parker Dynamo Models}
\label{parker} We first consider a simple dynamo model, a
straightforward generalization  of the initial Parker (1955)
migratory dynamo. Following Parker, we average the mean-field
equation with respect to the radial coordinate $r$ to arrive at the
following set of equations (see {\it e.g.} Baliunas {\it et al.}, 2006)

\begin{equation}
\label{eq1} {{\partial B} \over {\partial t}} = Dg \sin \theta
{{\partial A} \over {\partial \theta}} + {{\partial^2 B} \over
{\partial \theta^2}} - \mu^2 B, \label{eqb}
\end{equation}
\begin{equation}
\label{eq2} {{\partial A} \over {\partial t}} = \alpha B +
{{\partial^2 A} \over {\partial \theta^2}} - \mu^2 A \label{eqa}
\end{equation}
A formal derivation of the above equations and detailed
description of the non-dimensionalization is given in, {\it e.g.},
Sokoloff {\it et al.} (1996). Briefly, we use units of radius $R$ and global
diffusion time $R^2/\eta$, where $\eta$ is the magnetic diffusivity. Here
$B(\theta)$ represents the toroidal magnetic field and $A(\theta)$
is the azimuthal component of the vector potential for the poloidal
field. $\theta$ is co-latitude (so $\theta = 0$ corresponds to the
north pole). $g=g(\theta)$ is the radial shear of the differential
rotation. We cannot include a representation of a realistic solar
rotation curve in our simple 1D model, and choose for the sake of
simplicity $g =1$. $D$ is the dynamo number, which
incorporates the intensity of both sources of generation, the
alpha-effect and differential rotation, so the equations are given
in nondimensional form. The terms with second derivatives are
responsible for the latitudinal diffusion of magnetic field.  The
relaxation term proportional to $\mu^2$ represents the radial
diffusion of magnetic field (see for details {\it e.g.} Kuzanyan and
Sokoloff, 1996). We choose $\mu =3$, which corresponds to a
convective shell occupying $1/3$ of the solar radius.
Equantions~(\ref{eq1}) and (\ref{eq2}) are solved by timestepping on a finite
difference grid with $N$ points distributed uniformly in
$0\le\theta\le \pi$; our standard resolution is $N=101$.

We consider a simple algebraic quenching of $\alpha$-effect in the
form

\begin{equation}
\alpha = {{\alpha_0} \over {1 + B^2/B^2_0}},
 \label{quench}
 \end{equation}
where $\alpha_0$ is a nominal, unquenched value of the alpha-effect
and  $B_0$ is a field strength at which dynamo action is stabilized
by nonlinear effects, say the equipartition strength.

We appreciate that the actual form of dynamo quenching for solar
dynamo can be much more sophisticated that the simple illustrative
form~(\ref{quench}) (cf. {\it e.g.} Gruzinov and Diamond, 1994). In
particular, arguments of magnetic helicity conservation lead to a
scheme with a differential equation for $\alpha$ (Zhang {\it et al.}, 2006)
which seems to give additional options to produce stochastic behavior of
the dynamo generated magnetic field. Here, however, we restrict
ourself to the simple scheme~(\ref{quench}).

We include $\alpha$-quenching by the toroidal field only,
taking into account that the toroidal field is usually much larger
than the poloidal. We keep the standard latitudinal profile
$\alpha_0 \propto \cos \theta$ and impose fluctuations which
preserve the latitudinal profile in a given hemisphere by writing

\begin{equation}
\alpha = {{\cos \theta} \over {1+B^2}} (1 + r_i(t)).
 \label{fluct}
 \end{equation}
Thus we measure magnetic field in units of $B_0$ and incorporate the
typical amplitude of the alpha-effect in the definition of dynamo
number. The $r_i$ are pseudo-random numbers, supplied by a NAG
Library routine; $i=1$ for the Northern (N) hemisphere, $i=2$ for
the Southern (S). These values can be chosen independently, so we
can model deviations in equatorial antisymmetry of the alpha-effect
due to fluctuations. The time dependence of $r_i(t)$ is taken as
a piece-wise constant function. The correlation time $T$ is chosen
to be of order the nominal 11-year period. The fluctuations are
supposed to be independent over different time ranges of length $T$,
and Gaussian with standard deviation $\sigma$. We exclude very
strong fluctuations, larger then $5 \sigma$. This basic model we
refer to as Model I.

Our choice for the noise means that we consider global fluctuations
of $\alpha$ on the temporal scales of order the cycle length and
spatial (latitudinal) scale of the whole solar hemisphere. We do not
include in our analysis variations of short time and latitudinal
extent, on the scales of turbulent vortices, which are obviously
important for the short-term dynamics of solar activity. The
presence of long-term variations in the alpha-coefficient has been
reported from analysis of direct numerical simulations by
Brandenburg and Sokoloff (2002) and Otmianowska-Mazur {\it et al.}
(2006).

We also considered more sophisticated models for the noise.

(i) Model II. We now introduce in the sequences of time intervals
fluctuations of $\alpha$  between intervals of random length in
which there are no fluctuations. We define $\mu_{\Delta T},
\sigma_{\Delta T}$ to be the mean and standard deviations of the
time intervals when there are zero perturbations to $\alpha_N,
\alpha_S$. These parameters are the same for N and S hemispheres.
Perturbations to $\alpha$ are  {\it on} for fixed time $t_c$ (as the
previous model). {\it Off} times are of duration $\delta_{N,S}$, where
$\delta_{N,S}$ are gaussian random variables with mean $\mu_{\Delta
T}$ and standard deviation $\sigma_{\Delta T}$.

Limits are imposed on $\delta_{N,S}$:

\noindent (a) if $\delta_{N,S}<0.20\mu_{\Delta T}$, then
$\delta_{N,S}= 0.20\mu_{\Delta T}$, to avoid negative/very close to
zero intervals -- important if $\mu_{\Delta T}$ and $\sigma_{\Delta
T}$ are of similar size. So the minimum off interval is
$0.2\mu_{\Delta T}$.

\noindent (b) if $\delta_{N,S}>\mu_{\Delta T}+3\sigma_{\Delta T}$,
then  $\delta_{N,S}=\mu_{\Delta T}+3\sigma_{\Delta T}$.

Quite arbitrarily,  we take  $\sigma_{\Delta T} \sim \mu_{\Delta T}$.

(ii) Model III. We introduce a memory in the sequence of fluctuation
of $\alpha$ as follows. Let $p_n$ be a sequence of independent
pseudo-random gaussian numbers as supplied by the library routine
(see above also). Then we introduce $r_1 = p_1, r_2=p_2$ and

\begin{equation}
r_n = {{p_n+r_{n-1}}\over {\sqrt 2}}.
 \label{randmem}
 \end{equation}
The sequence $r_n$ has zero mean and the same standard deviation as
$p_n$  (the factor $\sqrt 2$ gives the normalization required). The
memory time for the sequence $r_n$ is of order $T$.

(iii) Model IV. Now we implement a 3-level memory , by writing

\begin{equation}
p_n={{p_{n-2}+r_{n-1}+r_n} \over {\sqrt 3}}.
 \label{rrandmem}
 \end{equation}

We note in advance that these memory effects turn out  not to play a
crucial role in our analysis and discussion.

\subsection{2D models with feedback on the differential rotation}
\label{MP}

With the aim of exploring other possibilities, we also studied the
behaviour of  a 2D model in which the nonlinearity is the feedback
of the Lorentz force on the differential rotation, using the code of
Moss and Brooke (2000). In order to retain some of the spirit of the Parker
dynamo, we considered a relatively thin shell, $0.8\le r\le 1.0$
where $r$ is the fractional radius.

For Model V, the initial rotation law is
$\Omega\propto r$. We kept $N=101$ latitudinal grid points as for Models~I--IV, but used
only $11$ uniformly distributed radial points, in what might be
termed a 1.5D model. (Taking the inner radius of the shell at
$r=0.9$ made no significant difference to our results.)
The random perturbations were implemented as in Model~I, with
independent hemispheres. The standard dynamo parameters $C_\alpha,
C_\omega$ were chosen to give a modestly supercritical dynamo, with
unperturbed magnetic period of approximately $0.022$. We allowed the
magnetic Prandtl number to take values of $1.0$ and $0.01$.
Unperturbed solutions were singly periodic.

For Model VI the initial rotation law has
$\Omega\propto\sin^2\theta$,  in an attempt to capture the essence
of the weak radial dependence
of the solar rotation in the bulk of the upper half of the
convection zone. $C_\alpha=5, C_\omega=-10^4$, otherwise Model~VI
was the same as Model~V. One significant feature of Model~VI is that the
dynamo is steady near to marginal excitation. This can be explained
by rotation laws with $\Omega=\Omega(\theta)$
being known to give radial migration
of field patterns. However, here the restricted radial extent of the model
inhibits this migration, resulting in a steady field.
Notwithstanding this rather
unsatisfactory feature, we present some results, as potentially illustrative
of processes in the upper part of the solar CZ.

\section{Results}

We performed a number of numerical experiments. In most cases a
modified Parker dynamo, as described above -- {\it i.e.} as Models I-IV
above -- was used  with $D=-10^3, \mu=3$ (a run in which $D=-3 \times
10^3$ is indicated by an asterisk in Table~\ref{tabres}). Then the
dimensionless period of the field $B$  ({\it i.e.} the nominal "22 yr"
solar cycle period) is $\tau \approx 0.25$.

For all cases we performed "short" computations, over a
dimensionless  time $2500$ ({\it i.e.} approximately 10000 cycles, which
is an order of magnitude longer than the solar activity
 reconstructed over the Holocene). For a number of cases we
made longer runs, over intervals of length $25000$. We summarize the
results obtained with Models I-IV in Table~1.

\begin{figure}
\begin{center}
\includegraphics[width=0.45\textwidth]{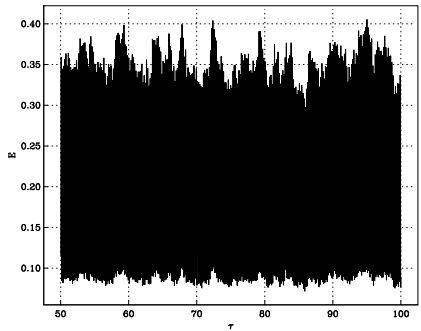}
\end{center}
\caption[]{\label{noGM} Time series of total energy for Model I, with $T=0.03$, $\sigma=0.05$.
This case does not show well defined Grand Minima-like
events. }
\end{figure}

\begin{figure}
\begin{center}
\includegraphics[width=0.45\textwidth]{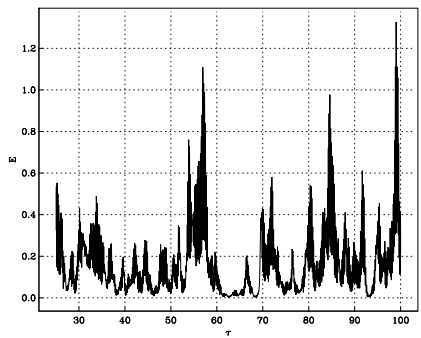}
\end{center}
\caption[]{\label{GM} Time series of total energy for Model III, with $T=0.15$, $\sigma=0.12$.
Here Grand Minima-like events can be identified. }
\end{figure}

We give in Figure~\ref{noGM} an example of a run where we were unable
to identify  clearly Grand Minima/Maxima (a - time series for
$\alpha$, b - time series for toroidal field at an  arbitrary point,
c - histogram for the cycle intensity), while Figure~\ref{GM} shows an
example of a run with pronounced events resembling Grand
Minima/Maxima.

For each model we consider $E$, the energy in the toroidal magnetic field.
First, the series are averaged over the nominal undisturbed cycle length
 (the analogue of the 11-year cycle) and passed through the Gleissberg
(12221) filter\footnote{The 12221 filtering is defined as $\langle x_i\rangle = 0.125(x_{i-2}+2x_{i-1}+2x_i+2x_{i+1}+x_{i+2})$. },
  which is often applied when studying long-term variations of solar activity in order to
 suppress the noise ({\it e.g.}, Gleissberg, 1944; Soon {\it et al.}, 1996),
 similarly to the original analysis of Usoskin {\it et al.} (2007).
Next, the distribution function of these averaged $E$ values was constructed
 and compared with that for sunspot activity (Figure~\ref{FD}).
Here we are interested mostly in the probability of occurrence
 of low values, and thus the robustness of the definition of Grand Minima.
Then we defined Grand Minima as periods when the cycle-averaged
 value of $E$ is systematically (for at least two consecutive cycles) below
 a threshold energy $E_{\rm th}$.
As discussed by Usoskin {\it et al.} (2007), Grand Minima of the solar activity
 are well-defined, due to the fact that the distribution function
 (Figure~\ref{FD}) is almost flat for SN below 10, implying an excess over a
 normal distribution.
The DFs of the model runs were visually compared to the "real" one and
 placed into one of the five
 categories (see column 5 in Table~1):
 "OK", if the DF is similar to that of the solar SN;
 "+", if there is a clear excess of low values (i.e the
simulated activity is dominated by the Grand Minima,
 and thus the statistics of their occurrence cannot be clearly determined) --
"++" denotes a stronger excess; and
 "-" that the low values are underrepresented compared to
that of the solar record. For some parameters the DF is bimodal and
this is indicated in the corresponding column of Table~1.  We note
that the identification of Grand Minima is relatively robust for
 the DF similar to the "real" one (denoted as "OK") in the Table,
 whereas it depends on the choice of $E_{\rm th}$ for both "+" and
 "-" DFs.
In our model, Grand Minima are not a special state of the dynamo,
but rather the result of
random fluctuations, and thus they cannot be unambiguously identified.

Next, the cumulative probability WTD\footnote{The waiting time is
defined between centres of successive Grand Minima.} was
constructed, in a similar way to that shown in Figure~\ref{WTD}, and
the shape of its tail was assessed as being either close to an
exponential, or deviating from it. This is indicated in column 6  of
the Tables. If a significant deviation from an exponential tail was
observed, the same model was rerun with new randomization and a
10-times longer realization in order to improve the statistics. In
all such cases the WTD returns to a nearly perfect exponential form
when the enhanced statistics are taken into account. In other cases
(notably Model~VI) a non-exponential tail may appear for a limited
range of threshold values $E_{\rm th}$, but disappears for larger
and smaller values. All such cases (deviation from exponential WTD
tail with normal statistics that disappears with enhanced statistics
or fragility with respect to choice of $E_{\rm th}$) are denoted by
"exp*" in Tables~1 and 2. We stress that the entries in the WTD
column refer to the tail of the distribution of the waiting time
between successive Grand Minima, whereas the DF entries refer to the
shape of the histogram  of SN ($E$ values) at low activity level.

\begin{table}
\label{tabres} \caption{Results for selected cases of Models~I-IV.
Model parameters are:  $n$ - run number (an asterisked run number
run means that $D=-3\times 10^3$ instead of usual $D= - 10^3$),
$\iota=1$ means that the $\alpha$-fluctuation is the same in both
hemisphere while $\iota = 2$ means that the fluctuations in both
hemisphere are independent, $\sigma$ is standard deviation and $T$
the mean of the random "off" times (for models I, II, IV the "off"
times are constant, and so $T$ is this constant value). Columns
under "Results":  DF gives the nature of the Distribution function
(excess/lack of low values denoted as "+"/"-", "++" denotes a very
strong excess,  bimodal stands for bimodal DF);  WTD describes the
shape of WTD ("exp" stands for the clear exponential tail, while
"exp*" denotes cases when a deviation from the exponent is observed
in some parameter range with lower statistucs but becomes
exponential with higher statistics.) In run 38$^\dagger$ we also
tried a different long realization of random fluctuations with the same
 governing parameters as in the run 38. See the text for
further details. }
\begin{tabular}{llllcl}
\hline
 \multicolumn{4}{c}{Model parameters}&\multicolumn{2}{c}{Results}\\
 \hline
 $n$ & $\iota$  & $\sigma$  &$T$ & DF  & WTD \\
 \hline
 \multicolumn{6}{c}{Model I}\\
 \hline
          2       & 1      &   0.2 & 0.12  & -  & exp \\
          3       & 1      &   0.4 &  0.12 &  OK & exp  \\
          4      & 1      &    0.6  & 0.12  & OK & exp \\
          5     &  2 &    0.2  & 0.12  & OK  & exp \\
          6     &  2 &     0.4 &  0.12  & + & exp \\
          7    &  2 &  0.2 &  0.06 & -  & exp   \\
        8*      & 2 &  0.2  & 0.20  & bimodal & exp*\\
         9      &    2  &   0.2 &  0.12  & bimodal & exp\\
         10     &    2 &    0.3  & 0.12 & + & exp \\
        11     &    2 & 0.25 & 0.12 & OK &  exp*  \\
        12     &    2 & 0.25 & 0.06 & - & exp* \\
        13     &    2 &  0.25 & 0.24  & + & exp \\
        \hline
 \multicolumn{6}{c}{Model II}\\
 \hline
21         & 2  &  0.25 & 0.12  & + & exp \\
22         & 2   & 0.30  &0.12  & + & exp   \\
  \hline
 \multicolumn{6}{c}{Model III}\\
 \hline
31       &    2   & 0.25 & 0.12  & + & exp \\
32       &    2   & 0.25 & 0.20  & ++ & exp \\
33       &    2   & 0.25 & 0.06  & ++ & exp \\
34       & 2   & 0.12 &  0.06 & - & exp \\
 35      &    2   & 0.12 & 0.20  & + & exp  \\
36       &    2   & 0.15 & 0.12  & OK  & exp*  \\
37       &    2   & 0.15 & 0.06  & OK & exp* \\
38      &    2   & 0.15 & 0.20  & OK  & exp* \\
38$^\dagger$      &    2   & 0.15 & 0.20  & OK & exp* \\
 \hline
 \multicolumn{6}{c}{Model IV}\\
 \hline
41       &      2  &  0.15  & 0.12  & OK & exp* \\
42       &      2  & 0.10   & 0.12  & - & exp \\
43       &      2  & 0.15   & 0.20  & OK & exp* \\
\hline
\end{tabular}\\
\end{table}

\begin{table}
\label{tabres2}
\caption{Results for selected cases for Models IV and VI.
Notation is as for Table~1, except that $Pr$, the magnetic Prandtl number, is
also tabulated; the label "exp*" is explained in the text.}
\begin{tabular}{lllllcl}
\hline
 \multicolumn{5}{c}{Model parameters}&\multicolumn{2}{c}{Results}\\
 \hline
 $n$ & $i$  & $\sigma$  &$T$ &$Pr$ & DF  & WTD\\
 \hline
 \multicolumn{6}{c}{Model V}\\
 \hline
          61      & 2      &   0.2 & 0.05 &1.0  & - & exp \\
          62      & 2      &   0.4 & 0.005  &1.0&  OK & exp  \\
          63     & 2      &    0.6  & 0.005 &0.01  & OK & exp* \\
        \hline
 \multicolumn{6}{c}{Model VI}\\
 \hline
71         & 2  &  0.20 & 0.005 & 1.0  & - & exp \\
72         & 2   & 0.20  &0.005 & 0.01 & - & exp*   \\
73         & 2   & 0.20  &0.010 & 1.0 & - & exp   \\
74         & 2   & 0.40  &0.010 & 1.0  & - & exp   \\
75         & 2   & 0.40  &0.010 & 0.01 & ++ & exp*   \\
76         & 2   & 0.20  &0.010 & 0.01 & - & exp   \\
\hline
\end{tabular}\\
\end{table}

The general result of this investigation is that the simple model
under consideration reproduces in a suitable parameter range many
features of the sequence of Grand Minima, as deduced from the
available observations. It means that we can in a more or less clear
way separate the epoch of normal cycles and Grand Minima, and that
the sequence of Grand Minima looks random rather than
periodic. We note again that the model considered does not contain
any specific mechanism to produce Grand Minima,  rather  that the
Grand Minima occur as a result of random fluctuations of the
parameters governing the dynamo. We appreciate that the model does
not reproduce the available phenomenology completely, {\it i.e.} we were
unable to reproduce the non-Poissonic tails of the WTD of the
isotopic record.

We have also investigated to what extent the results are robust when a more
complicated semi-2D model is considered. Table~2 gives results for
Models V and VI -- see also the discussion in Sect~\ref{disc}. Of
course, there are many possible ways to make the model more
realistic and it is practically impossible to search the parameter
space in full detail.  However, at least at first sight, the general
message seems to be clear: for the more complicated models we do not
see anything basically new compared with the simpler one.

\section{Long-term Dynamics of Solar Activity}

Let us summarize the results obtained. We have tried to simulate the
occurrence of solar Grand Minima, as being an effect of fluctuations
in the governing parameters in a simple model of the solar dynamo,
{\it i.e.} the Parker migratory dynamo. The model demonstrates some
aspects of this phenomenon within a suitable range of the dynamo
parameters. The sequence of simulated Grand Minima appears chaotic
and in this sense is similar to the observational data. The
statistics of Grand Minima occurrence looks however different from
that reconstructed from observations. While the WTD of observed
Minima demonstrates a substantial deviation
 from the exponential tail, the WTD of the simulated results
is nearly Poissonic.

We see that the effect of fluctuations is quite robust in the sense that
we do not obtain non-exponential tails in the probability
distribution of waiting times between Grand Minima/Maxima in any of
our simulations. Of course, the details of  the sequence of Grand
Minima/Maxima are model dependent. We even tried to modify the idea
under discussion quite substantially  by replacing the
alpha-quenching nonlinearity by feedback of the Lorentz force on the
rotation law. In all these cases we still obtained Poisson-like
distributions of Grand Maxima/Minima (or no such events at all).
However,  our investigations are largely limited to 1D models ({\it i.e.}
models which can be considered as a direct generalization of the
original Parker scheme), except for some experiments with the
Lorentz force nonlinearity, where a nominal radial dependence was
introduced in what might be termed a "1.5D" model.

Further elaboration of the dynamo model to fully 2D form
necessitates  substantially more complicated numerics, and our
ability to clarify the situation in detail is limited. We did
demonstrate however by one example that even fully 2D models with
Lorentz force feedback and on an underlying solar-like rotation law give a
Poisson-like sequence of Grand Minima/Maxima.  On the other hand,
there still remain some hopes that more complicated dynamo models
({\it e.g.} Moss and Brooke, 2000; Brooke {\it et al.}, 2002) can provide examples
of determinent intermittency, {\it i.e.} highly non-Poissonic behaviour of
the Grand Minima/Maxima sequence, for certain parameter choices,
notably a small value of the magnetic Prandtl number.

We believe that the disagreement between statistics of model results
and observations is plausibly related to the limited length of the
observational record,
 which is not long enough to be considered as statistically stationary in time.
On the other hand, the time series of numerical models can be made long enough
 to be statistically stationary.
When the length of a simulated data series is comparable with the length of
observational data, a significant deviation from the exponential WTD, similar to
 the observations, is found in several simulations.
However, this deviation disappears in all analyzed runs after extension of the
 simulated series (such cases are denoted as "exp*" in Tables~1 and 2.

\begin{figure}
\vspace{1.0cm}
\includegraphics[width=0.75\textwidth]{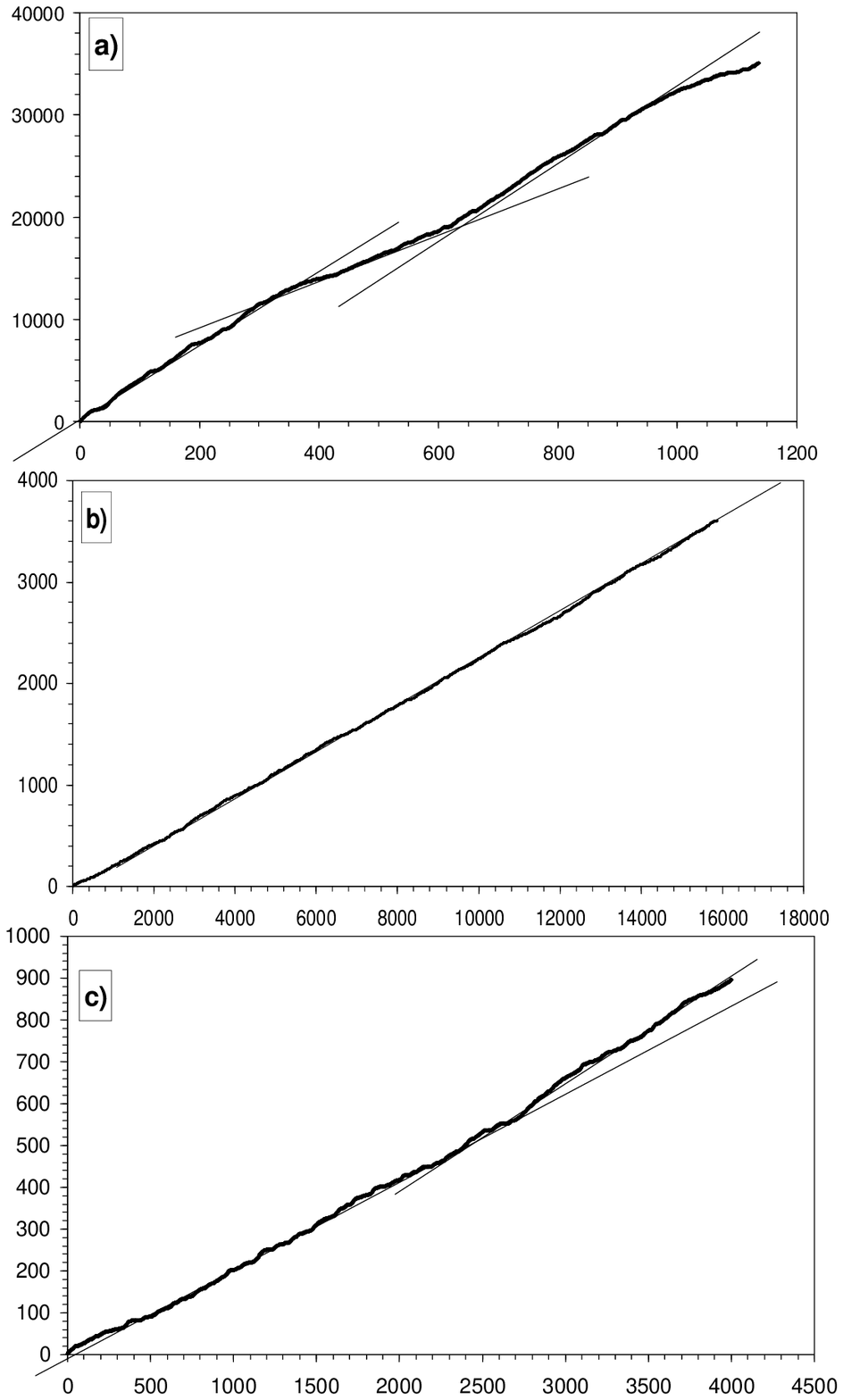}
 \caption{(a) Cumulative sunspot number against time, (b) run~41, full dataset,
cumulative energy vs time, (c) run~41, truncated dataset, cumulative
energy vs  time; we show strait lines which approximates the plots
presented}
\end{figure}


These differences can be related to the limited length of the series
which is not enough to reach statistical stationarity, as suggested
by the following test. Let us consider a time series $f_n = f(t_n)$
where the instants of observations $t_i= n \tau$. If the length of
the time series is sufficient to consider it as a stationary random
process then the random quantity $g_n = \Sigma_0^n f_i$ is expected
to behave as $g_n = \langle f\rangle n + h(n)$ where $\langle
f\rangle$ is the mean of the random process $f_n$ and $h(n)$ is a
function that grows more slowly then $n$. By plotting $g_n$ against
$n$ and comparing the results with a linear trend we can check
whether the time series is long enough to be consider as a
realization of a stationary random process. We performed this test
with the record of solar activity which has been exploited to
provide the sequence of Grand Minima (Figure~5a
here) and recognize there substantial deviations
from linear-law behaviour. We compare it with the simulated time
series for certain of the runs indicated as "exp*". We show in
Figure~5b the results obtained for the entire run 41 (of length 15873)
 -- it appears close
to linear. If however we restrict the length of this simulated
series to the first 4000 records the cumulative function
demonstrates substantial deviations from a linear behavior (Figure 5c).
Other runs marked as "exp*" in the
Tables 1 and 2 demonstrate similar results when this test is
applied.

 Note that the discussion above assumes implicitly that
solutions of the dynamo models with fluctuations in the dynamo
governing parameters can contain time scales much longer than the
cycle length and the average time between Grand Minima.
Justification of such assumptions needs specific numeric
simulations which are
 outside of the scope of this investigation.

We conclude that the observational data demonstrate similar
behaviour to the data with simulated data of the runs indicated as
"exp*" provided the comparable timescales are considered. We expect
that the observations of solar activity would give exponential
waiting time distribution for the sequence of Grand minima provided
such long record would be available.

\section{Discussion and Conclusions}
\label{disc}

We have demonstrated that the phenomenon of the occurrence of solar
Grand Minima can be simulated as an effect of fluctuations in the
governing parameters in a simple model of solar dynamo at least in
the framework of the interpretation of observation suggested. We
stress that the limited nature of the observations available does
not make it possible to compare the results of simulations
 and observations in
complete detail ; however we do not see in the observational data
anything that is basically incompatible with the simulations. Thus,
simulations in regimes marked as "exp*" in the tables
 look close to the observed phenomenology  and might be regarded
to be not inconsistent with the observations.

However, since the results cannot be directly compared in
 the statistical sense, other possibilities which exist
to explain the seeming disagreement between simulations and
observations deserve consideration. First of all, the phenomena of
Grand Minima and Maxima may be associated with fine details of the
solar dynamo (for example, the exact shape of the solar rotation
law), rather than being a general property of nonlinear dynamos in a
spherical shell for a suitable parameter range. (Our experiments
with a realistic rotation law in a 2D dynamo model, although not
encouraging, were too limited to rule this out completely.) A further
possibility is that the dynamo mechanism itself produces a
Poisson-like sequence of Grand Maxima/Minima, but there are also
long-term trends in solar hydrodynamics (on the scale of thousand
years) which affect the timescale of the weighting time and mimics
the non-Poisson behaviour. This could be, for example, via the
Reynolds stresses that drive the differential rotation. Another
option is that the non-Poissonic nature of the observed sequence of
Grand Minima/Maxima is an artefact of the limited statistics.

We stress once more that we present here a development of a very
strong suggestion that the solar dynamo engine does not contain any
specific mechanism that produces Grand Minima (and Maxima), but that
they are rather a result of random fluctuations in the dynamo
governing parameters. On one hand, the ability to convert random
noise into a sequence of clearly separated events looks  an
intriguing feature of the dynamo. This ability however can be
considered as an example of intermittency, which is a known property
of various nonlinear systems  where it can produce various spatial
or/and temporal structures from random noise (see {\it e.g.} Zeldovich {\it et al.}, 1991). On the other hand, it looks more than plausible that the
solar dynamo does possess something specific that allows fine tuning
of WTD of Grand Minima, which produces non-Poisson tails. However
we can not identify this feature of the dynamo engine at the moment,
and realize that the physical mechanism behind  the occurrence
of Grand Minima may be only partly or not at all related to random fluctuations
({\it e.g.} Petrovay, 2007).

We appreciate that the problem considered here belongs to the
general topic of the influence of
noise, which is addressed in many fundamental papers. Our ability to
exploit the deep methods suggested in this area ({\it e.g.}  Abarbanel {\it et al.}, 1993) is however restricted by the limited nature of the
available observational data. Note however that we have incorporated
some memory effects into models III, IV and V  so that, in principle
at least, we are going slightly beyond studying the effects of
random noise and the expected associated chaos. Also that it is not
a priori altogether obvious how these random inputs will appear
after passing through the dynamo "machine".

 Our work is based on the tacit assumption that the sunspot
number is linearly related to the magnetic energy in the dynamo.
This is a common asssumption in solar cycle modelling, but
it is quite possible that the number and size of the
active regions appearing on the
surface might be more plausibly taken to be proportional to the
toroidal flux, rather than to the energy. However this option
leads to similar conclusions to those presented above. On the
other hand, it also seems possible that there
could be a threshold effect at
play here, so that active regions only emerge if the toroidal field
strength exceeds some minimal value ({\it e.g.} Ruzmaikin, 2001). This would
introduce a marked nonlinearity into the relationship between
activity indices and toroidal field parameters. It is clear that the
results obtained can be sensitive to this nonlinearity which,
in principle, we could investigate.
Accordingly, our investigation of the Grand Minima
phenomenon has to be considered
to some extent as illustrative until the influence of the nonlinearity
is resolved.
A further possibility in a 2D model
is to use the toroidal flux or energy in the immediate sub-surface
region as a proxy for surface activity.

Finally, we mention the regime with so-called ``dynamo outbursts''
observed in the VKS dynamo experiment (Ravelet {\it et al.}, private communication) as presenting
one further topic that may be relevant to our investigation. We note
that episodes from the time series of magnetic field evolution taken
from a sensor in this experiment look very similar to that
presented in Figure~4.

We recognize that our results are not positive in the sense of
answering in a clear-cut manner the key question of whether a simple
model, such as we have considered, is able to reproduce the
observed statistics of the occurrence of solar Grand Minima. However
we do feel that we may have provided some insight into the question
of the statistical stability of the observations, and to have
provided information and guidance for future
investigations.

%

%

%
\begin{acks}

D.S. is grateful  to the  Royal Society for supporting his visit to
Manchester, and also to RFBR for financial support under grant
07-02-00127. D.M. and D.S. thank the Finnish Academy of Science and
Letters (V\"ais\"al\"a foundation) for supporting their visits to
the University of Oulu.

\end{acks}

%
%
%
%
%
%

\end{article}
\end{document}